\documentclass[aps,prc,preprint,superscriptaddress,nofootinbib]{revtex4-2}

\usepackage{amsmath}
\usepackage{graphicx}
\usepackage{units}
\usepackage{url}
\usepackage{ulem}

\usepackage[dvipsnames]{xcolor}
\usepackage{color}
\usepackage{tabularray}
\usepackage[export]{adjustbox}

\newcommand{\Ru}{${}^{96}_{44}$Ru\,}
\newcommand{\Zr}{${}^{96}_{40}$Zr\,}
\newcommand{\trento}{T$_{\rm R}$ENTo\,}

\usepackage{hyperref}

\begin{document}
\title{Impact of the  pre-equilibrium stage for the determination of nuclear geometry in high-energy isobar collisions}

\author{Fernando G. Gardim}
\affiliation{Instituto de Ci\^encia e Tecnologia, Universidade Federal de Alfenas, 37715-400 Po\c{c}os de Caldas-MG, Brazil}

\author{Andr\'e V. Giannini}
\email[]{andregiannini@ufgd.edu.br}
\affiliation{Faculdade de Ci\^encias Exatas e Tecnologia, Universidade Federal da Grande Dourados,
Caixa Postal 364, CEP 79804-970 Dourados, MS, Brazil}
\affiliation{Departamento de F\'isica, Universidade do Estado de Santa Catarina, 89219-710 Joinville, SC, Brazil}
\affiliation{Instituto de Ci\^encia e Tecnologia, Universidade Federal de Alfenas, 37715-400 Po\c{c}os de Caldas-MG, Brazil}

\author{Fr\'ed\'erique Grassi}
\affiliation{Instituto de F\'{i}sica, Universidade de S\~ao Paulo, Rua do Mat\~ao 
1371,  05508-090 S\~ao Paulo-SP, Brazil}

\author{Kevin P. Pala}
\affiliation{Instituto de F\'{i}sica, Universidade de S\~ao Paulo, Rua do Mat\~ao 
1371,  05508-090 S\~ao Paulo-SP, Brazil}

\author{Willian M.~Serenone}
\email[]{wmatioli@if.usp.br}
\affiliation{Instituto de F\'{i}sica, Universidade de S\~ao Paulo, Rua do Mat\~ao 
1371,  05508-090 S\~ao Paulo-SP, Brazil}
\affiliation{Department of Physics, University of Illinois at Urbana-Champaign, Urbana, IL 61801, USA}

\date{\today}
 
\begin{abstract}

Ultrarelativistic isobar collisions have been proposed as a useful tool to investigate nuclear structure. These systems are not created in equilibrium, rather undergo a pre-thermalization stage. In this stage, some of the initial structure information may be lost and additional  effects introduced. The objective of this paper is to study this possibility in the extreme case of a ``free-streaming'' pre-equilibrium stage.
We do this by computing estimators for ratios of various measured (or measurable) quantities (elliptic and triangular flows, mean transverse momentum and associated cumulants,  correlators between elliptic or triangular flows and mean transverse momentum, symmetric cumulant and two-plane correlator) and study their sensitivity to the duration of the free-streaming stage. 
We find that the correlators between elliptic or triangular flows and mean transverse momentum, the so-called $\rho_2$ and $\rho_3$, are indeed sensitive to the duration of the free-streaming stage and that the normalized symmetric cumulant, $\varepsilon NSC(2,3)$ might also depend on this duration.

\end{abstract}


\maketitle

\section{Introduction}

The successful application of the hydrodynamic framework to describe high energy heavy-ion collisions has opened a new door to study nuclear structure. Experimental data from the collaborations STAR \cite{STAR:2015mki, STAR:2021mii, Xu:2022ikx}, ALICE \cite{ALICE:2018lao, ALICE:2021gxt}, CMS \cite{CMS:2019cyz}  and ATLAS \cite{ATLAS:2019dct, ATLAS:2022dov} as well as theoretical studies \cite{Heinz:2004ir, Shou:2014eya, Xu:2017zcn, Li:2019kkh, Giacalone:2021udy, Giacalone:2021uhj, Jia:2021tzt, Jia:2021qyu, Jia:2021wbq, Zhang:2021kxj, Jia:2021oyt, Xu:2021uar, Bally:2021qys, Nijs:2021kvn, Zhao:2022uhl}, have indicated the importance of nuclear deformations and radial distributions of protons and neutrons in determining the measured multiplicity distribution and collective flow in these collisions.

Among deformed nuclei, the isobars \Ru+\Ru and \Zr+\Zr collisions at $\sqrt{ s_{NN}}=200$ GeV studied by the STAR Collaboration  \cite{STAR:2021mii,Xu:2022ikx} are of particular interest due to the similarities between the nuclei and the precision of the measurements. In these collisions, \Ru exhibits higher particle production (by up to 10 percent), larger elliptic flow (by up to 3 percent), and smaller triangular flow (by 7 percent for central collisions) compared to \Zr. These effects are relatively small, but still significant enough to be detected with high statistical precision and systematic uncertainties cancellation in ratios.

On the theory side, a large number of collision simulations is needed to achieve such high precision, and the process must be repeated when poorly understood nuclear geometry parameters are changed.  Hydrodynamic-based simulations are computationally expensive, so one approach is to use the known mapping between initial conditions and anisotropic flow \cite{Gardim:2011xv,Gardim:2014tya,Niemi:2012aj,Niemi15,Fu15}. Numerous studies have employed the aforementioned methodology, such as \cite{Jia:2021tzt, Jia:2021qyu, Jia:2021wbq, Zhang:2021kxj}. Nevertheless, a limited number of works have utilized hydrodynamic simulations \cite{Xu:2021uar, Nijs:2021kvn, Zhao:2022uhl}.

State-of-the-art simulations of heavy-ion collisions are commonly  based on a hybrid-chain of models, comprising four distinct stages: firstly, the generation of initial conditions; secondly, a pre-thermalization stage which serves to connect the initial conditions to the onset of hydrodynamic expansion; thirdly, the calculation of fluid expansion through the use of viscous relativistic hydrodynamics, and finally, a hadronic transport phase which occurs at the end of the hydrodynamic expansion. While the multiplicity distribution and collective flow are sensitive to the value of viscosity, as well as the inclusion of final state interactions and free streaming, it is expected that differences in the ratios of these observables only arise due to differences in nuclear structure. Although this assumption is reasonable, the precision of experimental data requires 
that it be tested. This has not be done systematically so far.
An indirect study of viscosity and hadronic cascade effects using a transport model was recently conducted in \cite{Zhang:2022fou}.

A fundamental physical question in the preceding hydrodynamic description is the time after which the viscous hydrodynamic description becomes valid. Early investigations indicated that to reproduce the observed large hadron momentum anisotropy, this time must be smaller than about 1 fm \cite{Heinz:2001xi}. However, studies in the strong and weak coupling limits have not fully explained this rapid thermalization, which triggered new approaches to understand when hydrodynamics does apply (see e.g.~the reviews \cite{Romatschke:2016hle,Florkowski:2017olj,Romatschke:2017ejr}). The objective of this paper is to investigate the effect of the pre-thermalization on the description of isobar collisions. This is done in the extreme scenario where immediately after production, partons are free streamed until the beginning of the hydrodynamic evolution. We note that such an approach has been used in various recent Bayesian analysis, where the most likely range of parameter values for the hydrodynamic scenario is sought \cite{Bernhard:2019bmu,JETSCAPE:2020shq,JETSCAPE:2020mzn,Parkkila:2021tqq, Nijs:2020roc,Nijs:2020ors,Liyanage:2023nds}.
We perform  our study using well known estimators for elliptic and triangular flows, but we also present estimators that have received little attention for isobars for quantities such as mean transverse momentum and associated cumulants, correlators between elliptic/triangular flow and mean transverse momentum, symmetric cumulant, mixed harmonic correlator.

\section{Setup of the simulations}

Albeit many sophisticated models for nuclei exist from low-energy studies, the most popular way to model the nucleus in a ultra-relativistic collision is by sampling a Woods-Saxon distribution. The original distribution is spherically symmetric, hence not suitable for studying nuclei which are expected to have a deformed shape, such as \Zr and \Ru. The non-spherical shape is then taken into account by introducing an angular dependence in the nuclei radius, such as  done in \cite{Jia:2021tzt, Giacalone:2021uhj, Bally:2022vgo}. Hence, the probability $P(r,\, \theta,\, \varphi)$ of finding a nucleon in the position
$\vec{r} = (r,\, \theta,\, \varphi)$ is proportional to the nuclear density profile
\onecolumngrid
\begin{align}
\begin{aligned}
\rho(r,\, \theta,\, \varphi)  & =  \frac{\rho_0}{1+\exp\left\{[r-\mathcal{R}(\theta,\, \varphi)]/a\right\}} \\
\mathcal{R}(\theta,\, \varphi) & = R_0\left\{1+\beta_2 \left[Y^0_2\left(\theta,\, \varphi\right)\cos\gamma + \frac{2}{\sqrt{2}}\sin \gamma\, \mathrm{Re} Y^2_2\left(\theta,\, \varphi\right) \right]+\beta_3Y^0_3\left(\theta,\, \varphi\right) \right\}\,,
\label{eq:WSdeform}
\end{aligned}
\end{align} 
where the nuclear radius $R_0$, diffuseness $a$, quadrupole deformation parameter $\beta_2$, triaxiality deformation parameter $\gamma$ and octupole deformation parameter $\beta_3$ characterize the nuclei structure. It is expected that  variations in these parameters are reflected in final-state observables. Here,  motivated by the values for these parameters present in \cite{KIBEDI:2002wxc,Pritychenko:2013gwa,Zhang:2021kxj} for \Ru and \Zr, we sample 6 distinct configurations, where we progressively tune  our parameters from a triaxial \Ru nucleus (configuration 1) to the \Zr nucleus (configuration 6). The parameters used are shown in Table \ref{tab:nuclear_param}.
\begin{table}
\centering
\caption{Nuclear structure parameters employed for generating initial conditions of isobar collisions.}
\label{tab:nuclear_param}
\begin{tabular}{|c|c|c|c|c|c|c|}
\hline
\textbf{Configuration Label} & $\mathbf{R_0}$ \textbf{(fm)} & $\mathbf{a}$ \textbf{(fm)} & $\mathbf{\hspace{.2cm} \beta_2 \hspace{.2cm}}$ & $\mathbf{\hspace{.2cm}\beta_3\hspace{.2cm}}$ & $\mathbf{\gamma}$  \\ 
\hline
Configuration 1 & 5.09 & 0.46 & 0.16 & 0 & $\pi/6$ \\
\hline
Configuration 2 & 5.09 & 0.46 & 0.16 & 0 & 0 \\
\hline
Configuration 3 & 5.09 & 0.46 & 0.16 & 0.20 & 0 \\
\hline
Configuration 4 & 5.09 & 0.46 & 0.06 & 0.20 & 0 \\
\hline
Configuration 5 & 5.09 & 0.52 & 0.06 & 0.20 & 0 \\
\hline
Configuration 6 & 5.02 & 0.52 & 0.06 & 0.20 & 0 \\
\hline
\end{tabular}
\end{table}

\begin{figure}[hbtp]
\begin{center}
  \includegraphics[width=\linewidth]
  {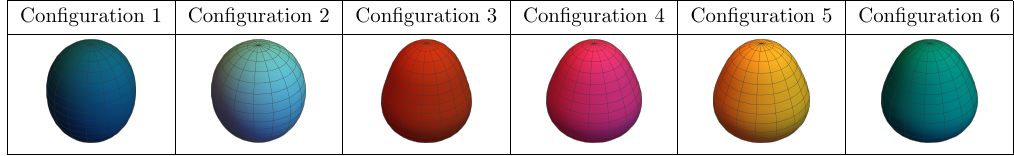}
   \caption{Shapes of different configurations in the nuclei, using the parameters of Table \ref{tab:nuclear_param} in Eq. \eqref{eq:WSdeform}.}
\label{fig:shape}
  \end{center}
\end{figure}

In Table \ref{tab:nuclear_param}, when going from one line to the next, only one parameter is changed. Configurations 1 and 2 are similar except that the former is a triaxial ellipsoid and the later is a spheroid. An octupole deformation (pear shape) has been introduced when going from configuration 2 to 3. When going from case 3 to 4, the quadrupole deformation (ellipsoid shape) is reduced. Configurations 4 and 5 are similar except that the later has a  larger diffuseness. Lastly, the nuclear radius is reduced when going from configuration 5 to 6. These systematic modifications in the nuclear structure parameters allow to investigate eventual changes in the results when gradually moving from what is typically considered a Ruthenium nucleus (configuration 2) to what typically is used as a Zirconium nucleus (configuration 6). The effect of these different parameters in the shape of the nucleus can be seen in Fig. \ref{fig:shape}. Effects related to ``neutron-skin'' have not been included in our simulations, therefore, protons and neutrons follow the same nuclear distribution. The reader may refer to \cite{Xu:2021vpn} for a study on the neutron-skin effects.

Due to its flexibility and the ability to accept pre-generated (possibly deformed) nuclear configurations for projectile and target, the \trento model \cite{Moreland:2014oya} has been used to generate initial conditions for our simulations. Two nuclear configurations (\ref{eq:WSdeform}) are randomly selected from a set of $50 000$ pre-generated options and subsequently randomized in orientation. The Monte-Carlo Glauber model, where a value for the nucleon-nucleon cross-section $\sigma_{\rm NN}$ is needed, is then applied to identify which nucleons will participate in the collision, and to each participating nucleon, a density function $\rho_i(x,y,z)$ is assigned, namely a Gaussian function with standard deviation given by the {\it nucleon width} $w$. Each nucleon density is allowed to fluctuate according to a Gamma distribution with shape parameter $k$. By summing these densities and integrating over the $z$ direction for each nucleus, the {\it participant} thickness $T_{A,B}$ is obtained. These thicknesses are then used to calculate the energy density $\epsilon(x,y,\eta_s=0)$ as a {\it reduced} 
thickness between the two participant thicknesses given by
\begin{align}
\bar{\epsilon}(x,y) \equiv \lim_{\tau \to 0^+} \tau \epsilon(x,y,\eta_s=0) = N \left[\frac{T_A^p(x,y)+T_B^p(x,y)}{2}\right]^{1/p}\,,
\label{eq:eps}
\end{align}
where $N$ is a normalization constant and $p$ is called the {\it reduced thickness parameter}. Since in this work we will be dealing solely with ratios of quantities such as eccentricities and average radius of the system, the exact value of the normalization should not matter. However, in order to actually produce initial conditions with \trento, one needs to specify a value for its physical parameters ($N$, $w$, $\sigma_{\rm NN}$, $k$, $p$). In our simulations, these parameters were chosen to be the same as in~\cite{JETSCAPE:2020mzn} (because their analysis was done for both RHIC and LHC energies) and are outlined in Table \ref{tab:trento_pars} for completeness.
\begin{table}[]
    \caption{\trento parameters used to generate initial conditions\cite{JETSCAPE:2020mzn}}
    \centering
    \begin{tabular}{|c|c|c|} \hline
        \textbf{Parameter}                    & \textbf{Symbol}            & \textbf{Value} \\ \hline
        Normalization (GeV/fm$^2$)   & $N$               & 5.73  \\
        Nucleon width (fm)           & $w$               & 1.12  \\
        Cross-section (fm$^2$)       & $\sigma_{\rm NN}$ & 4.23  \\
        Fluctuation                  & $k$               & 0.907 \\
        Reduced thickness parameter  & $p$               & 0.063  \\\hline
    \end{tabular}
    \label{tab:trento_pars}
\end{table}

After the collision takes place, one needs to specify how long the produced system will stay in the pre-thermalization stage. Given the model chosen in this work, such characteristic is controlled by the free-streaming time, $\tau_{FS}$. It is worth noting that at the moment all assumptions related to the modeling of pre-equilibrium stage are \textit{ad-hoc} and such a quantity is currently chosen to have either a  constant value or is allowed to carry a dependence with the collision centrality. 

Regardless of the case, the value of $\tau_{FS}$ or its behavior with the collision centrality started to be determined (together with many other parameters needed in a complete hybrid hydrodynamic simulation) via Bayesian inference techniques~\cite{Moreland:2018gsh, Bernhard:2019bmu, JETSCAPE:2020mzn, JETSCAPE:2020shq, Nijs:2020roc, Parkkila:2021tqq, Nijs:2021clz, Parkkila:2021yha}. This does not mean, however, that the main characteristics of this stage are well understood. In fact, it depends on the many particularities of each statistical analysis as different collaborations have obtained values for $\tau_{FS}$ that vary more than 50\%: for instance, \cite{Parkkila:2021tqq} extracted $\tau_{FS}=0.901$ fm/c by considering Pb+Pb collisions at 5.02 TeV; on the other hand, \cite{Parkkila:2021yha} reported $\tau_{FS}=0.71$ fm/c improving on the previous study by considering a simultaneous analysis of Pb+Pb data at 2.76 TeV and 5.02 TeV. When allowing the free-streaming to change with centrality as in \cite{JETSCAPE:2020mzn}, with parameters from the Grad viscous correction model, for Au+Au  collisions at 200 GeV, $1.36 \lesssim \tau_{FS}\, ({\rm fm/c}) \lesssim 1.51$, a variation of almost 50\% when compared to the improved estimate from \cite{Parkkila:2021yha}. Lastly, a simultaneous analysis of p+Pb and Pb+Pb data at 5.02 TeV presented in \cite{Moreland:2018gsh} yielded $\tau_{FS} = 0.37$ fm/c. While a smaller free-streaming time is expected for p+A collisions due to the system size after the collision takes place, this estimate (which is also applied to Pb+Pb collisions if one makes use of the model outlined in~\cite{Moreland:2018gsh}) is $\approx 0.59\%$ smaller than the value quoted in~\cite{Parkkila:2021tqq} which only considered Pb+Pb data at the same collision energy.

\begin{figure}[hbtp]
\begin{center}
  \includegraphics[width=0.80\linewidth]
  {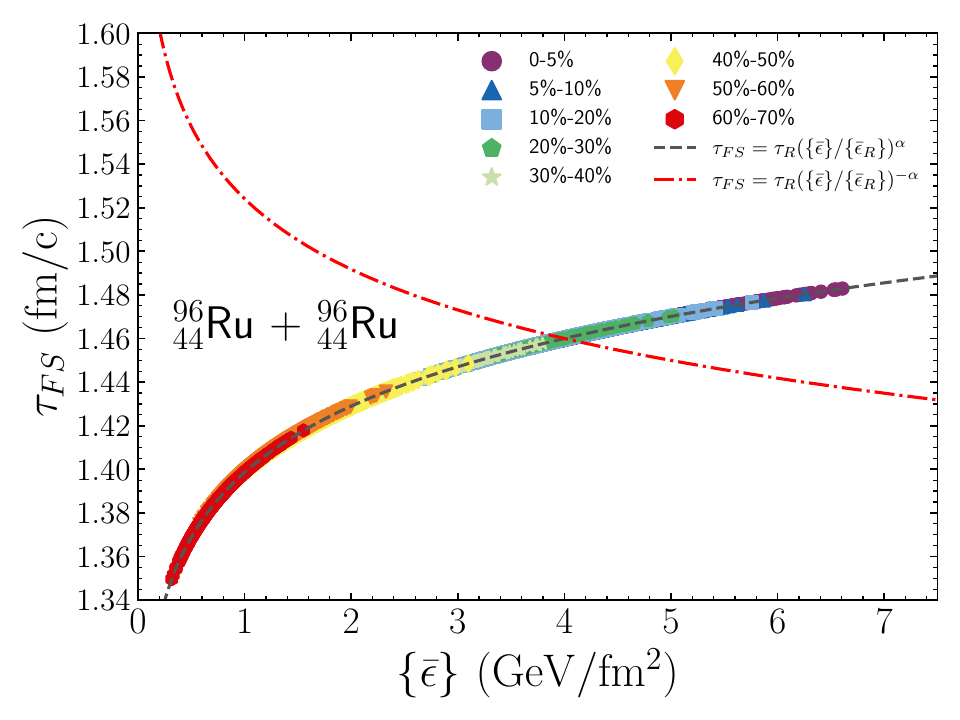}
   \caption{Free-streaming time as function of $\{{\bar\epsilon}\}$ in a mid-rapidity slice with $\tau_R = 1.46$ fm/c, $\bar\epsilon_R = 4$ GeV/fm$^2$ and $\alpha = 0.031$ \cite{JETSCAPE:2020mzn}. Centrality windows are also shown (see text). Very similar results are obtained for ${_{40}^{96}Zr} + {_{40}^{96}Zr}$. The $\alpha=-0.031$ case is shown by the red dot-dashed line. Both positive and negative $\alpha$'s  lead to comparable values of $\tau_{FS}$  for  central collisions.}
\label{fig:taufs}
  \end{center}
\end{figure}

Given the current lack of a detailed understanding of the pre-thermalization stage and the large uncertainty in the extracted values for $\tau_{FS}$, in the present study we consider three different choices for the free-streaming time: (\textit{i}) $\tau_{FS} = 0.4$ fm/c, (\textit{ii}) $\tau_{FS} = 1.0$ fm/c and (\textit{iii}) a centrality-dependent free-streaming time as implemented in \cite{JETSCAPE:2020mzn}: $\tau_{FS} = \tau_R(\{\bar{\epsilon}\}/\bar\epsilon_R)^{\alpha}$, where $\{{\bar\epsilon}\}$ denotes a (particular) average initial energy density in the transverse plane for a given collision, $\bar\epsilon_R= 4.0$ GeV/fm$^2$, $\tau_R = 1.46$ fm/c and $\alpha = 0.031$ are constants that were adjusted by comparison with experimental data. These choices cover the range of values determined by recent Bayesian analysis and allow for a systematic investigation of eventual impacts that the pre-thermalization stage may have on quantities related to the system shape. In Fig.~\ref{fig:taufs}, the free-streaming time for the centrality-dependent case is shown as function of $\{{\bar\epsilon}\}$ for Ru (very similar results hold for Zr). Centrality windows were estimated from ordering the total energy $E$ (integral on the transverse plane of the energy density in eq. (\ref{eq:eps}), 
$E=N \int d^2 x_\perp \left[\frac{T_A^p(x,y)+T_B^p(x,y)}{2}\right]^{1/p}$)
of 1 million \trento initial conditions in descending order for each colliding system. One sees that the free-streaming time varies between $\approx$ 1.35 and $\approx$ 1.49 fm in such a case. 

In Fig.~\ref{fig:taufs}, $\tau_{FS}$, the time when free-streaming stops and hydrodynamics is switched on, increases for larger energy densities. Though this is discussed and some arguments for this put forward in 
\cite{JETSCAPE:2020mzn}, one might expect the opposite: for larger densities, more collisions happen and hydrodynamics starts earlier. Therefore we have also included a curve that corresponds to a negative value for $\alpha$. For  central collisions, which are usually considered more interesting for isobar collisions, both positive and negative
$\alpha$ lead to similar fairly constant values. However we will discuss both cases when necessary below.

\section{Interplay between free-streaming and nuclear geometry}

Due to free-streaming, small scale features are expected to be smeared out and the uniform expansion results in decreased spatial asymmetries, i.e. smaller eccentricities $\varepsilon_n$, at the onset of the hydrodynamic phase, defined as 
\begin{align}
    \label{eq:ecc}
    \varepsilon_n & = \frac{|\{r^ne^{in\phi}\}|}{\{r^n\}}\,,
\end{align}
where $r$ and $\phi$ are the radial and azimuthal coordinates of the transverse $x-y$ plane in the center of mass reference frame. The $\{\cdots\}$ represents a (usual) energy density-weighted spatial average, differently from the arithmetic mean $\langle \cdots \rangle$, used here to denote average over events. Fluctuations of the eccentricity values can also be studied by computing the second and fourth order cumulants
\begin{align}
    \varepsilon_n\{2\} & = \sqrt{\langle \varepsilon_n^2 \rangle}\,, \\
    \varepsilon_n\{4\} & = \sqrt[4]{2\langle \varepsilon_n^2\rangle^2-\langle \varepsilon_n^4 \rangle}  \,.  
\end{align}

Eccentricities have long been expected to be good estimators for anisotropic flow coefficients
\cite{Gardim:2011xv,Gardim:2014tya,Niemi:2012aj}.  It has been argued that when there is a free-streaming stage and  a Laudau matching is done to switch to a hydrodynamic description, an initial momentum anisotropy  is created and 
could also play a role \cite{Broniowski:2008qk, Liu:2015nwa}. 
More recent papers indicate that for nuclear collisions (and some pre-equilibrium stage),  initial spatial
eccentricity is a good predictor for elliptic flow (see e.g. \cite{Sousa:2024msh, Giacalone:2020dln}). Therefore in this paper, we concentrate on spatial eccentricities.

\begin{figure}[htbp]
\begin{center}
  \includegraphics[width=.95\linewidth]{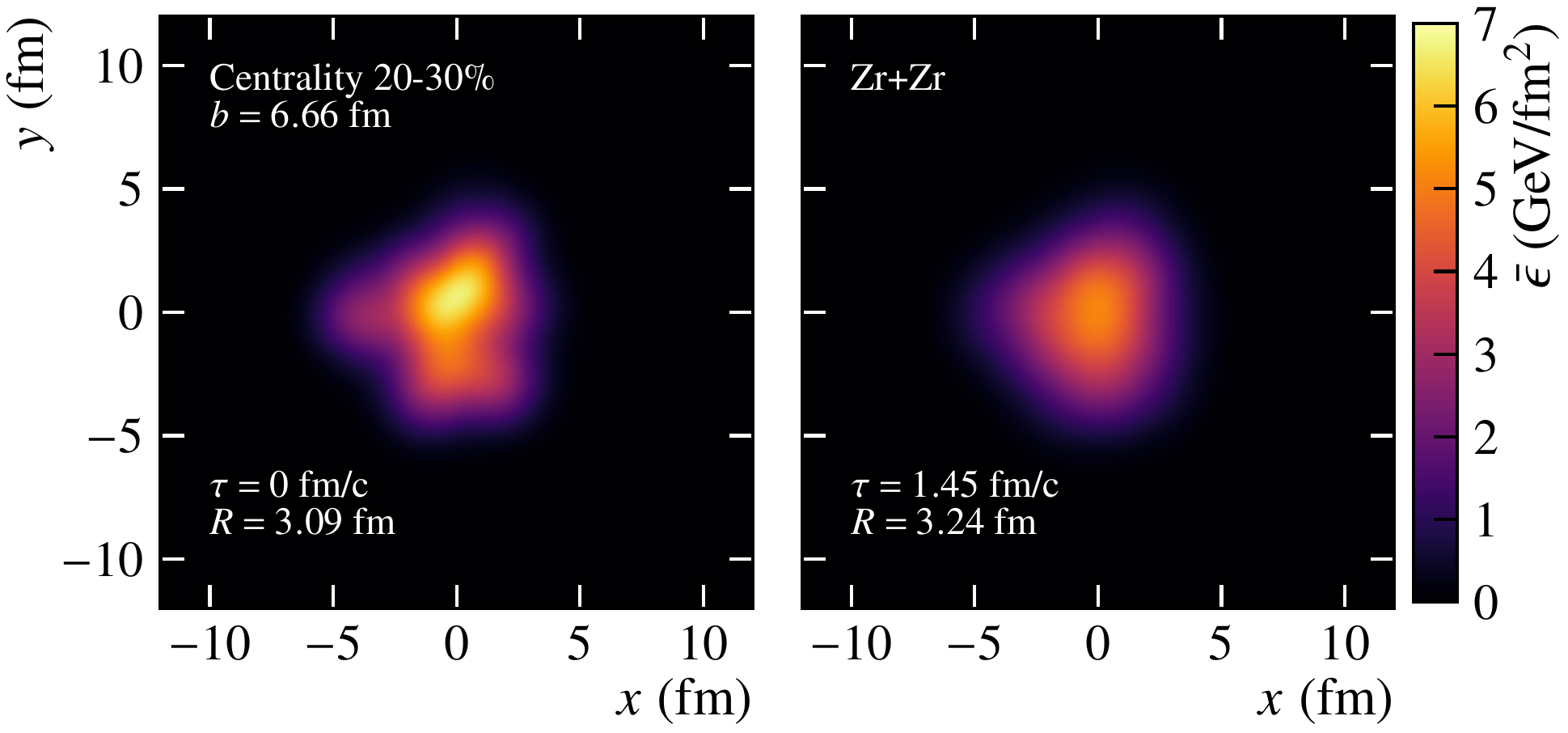}
  \caption{The mid-rapidity energy density slice of a  \Zr+\Zr event in the 20--30\% centrality window is shown  (left) initially 
  and (right) after free-streaming.}
\label{fig:ic}
  \end{center}
\end{figure}

Fig.~\ref{fig:ic} shows the same event at the initial time and at the end of the free-streaming (computed for configuration (iii)). One sees that the energy density distribution becomes more rounded, with a larger average radius $R\equiv \sqrt{\{r^2\}}$ as time flows. Therefore, due to  free-streaming, $R$ increases while the eccentricities decrease.

\begin{figure}[hbtp]
\begin{center}
  \includegraphics[width=.95\linewidth]{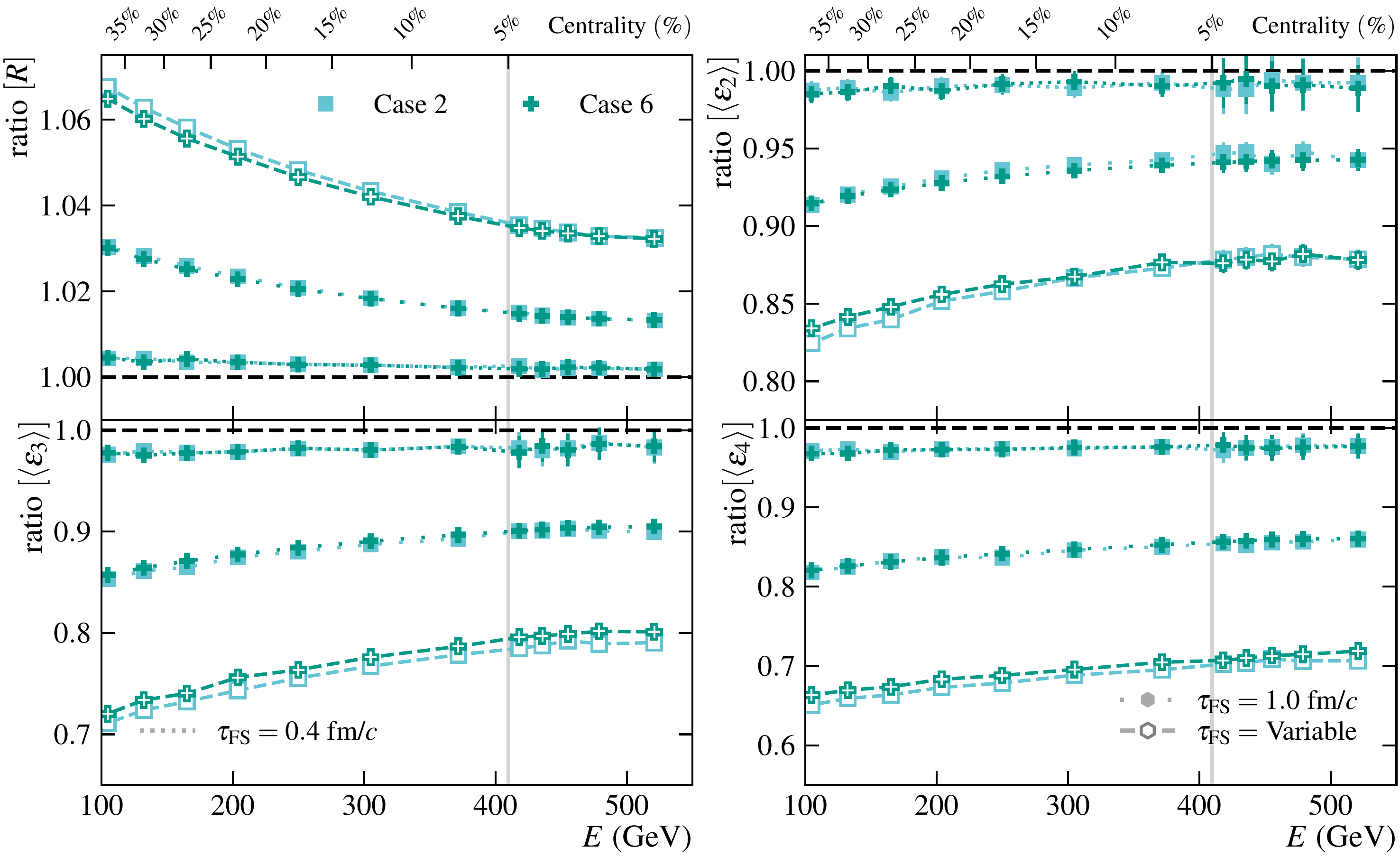}
   \caption{Root mean square radius $R$ and eccentricities $\varepsilon_n$ at the end of free-streaming stage divided by their initial values. Line styles correspond to different free-streaming times. One million minimum-bias events were used to generate this figure.
   Results for all configurations lie close so only those for configurations 2 and 6 are shown. 
   }
\label{fig:icfs}
  \end{center}
\end{figure}

This is shown more precisely in  Fig.~\ref{fig:icfs}. In this figure,  the ratio between a given  quantity computed after free-streaming and for the initial condition is used. It is a more effective mean of expressing the influence of the pre-thermalization stage. The following concise notation is used
\begin{align}
    {\rm ratio }[A] = \frac{\langle A \rangle_{FS}}{\langle A \rangle_{IC}}. \nonumber
\end{align}
Here and  below, to determine centrality bin edges, all generated events are used, regardless of the (geometrical) configuration they belong to in Table \ref{tab:trento_pars}. Such a procedure is different from the actual experiment, where centrality classification is based on the number of charged tracks together with information from Monte Carlo Glauber simulations for each collision system independently~\cite{STAR:2021mii}. Results are in general shown in bins of width 1\% in the 0--5\% centrality window and then in bins of 5\%; finer bins are used to $\rho_2$ and $\rho_3$, see below. When statistics demands, the 5 most central bins are merged in a single one of 0--5\%. Centrality bin edges can be read from Table   \ref{tab:centrbins} 
\begin{table}[htbp]
    \caption{Definition of the centrality bins as a function of E.}
    \centering
    \begin{tabular}{|c|c|c|c|}
        \hline
        \textbf{Centrality Bin} & \textbf{Energy Range (GeV)} & \textbf{Centrality Bin} & \textbf{Energy Range (GeV)}\\
        \hline
        0 -- 1\% & 700.00 -- 466.10 & 10-15\% & 275.69 -- 225.49 \\
        1 -- 2\% & 466.10 -- 444.98 & 15-20\% & 225.49 -- 183.31 \\
        2 -- 3\% & 444.98 -- 426.64 & 20-25\% & 183.31 -- 147.75 \\
        3 -- 4\% & 426.64 -- 409.73 & 25-30\% & 147.75 -- 117.71 \\
        4 -- 5\% & 409.73 -- 336.08 & 30-35\% & 117.71 -- 92.53  \\
        5 -- 10\% & 336.08 -- 275.69 &35-40\% & 92.53 -- 71.62 \\
        \hline
    \end{tabular}
    \label{tab:centrbins}
\end{table}
 
In Fig.~\ref{fig:icfs}, curves end up grouped by how long the produced system free-streamed: for eccentricities, one group of curves corresponds to $\tau_{FS} = 0.4$ fm/c, the middle one to $\tau_{FS} = 1.0$ fm/c and the last one to the centrality-dependent $\tau_{FS}$, estimated to be approximately between 1.35 fm/c and 1.49 fm/c from Fig.~\ref{fig:taufs}. Free-streaming effects are increasingly stronger (meaning, larger deviations from 1) going from $R$, to $\varepsilon_2 $, then  $\varepsilon_3 $ and finally  $\varepsilon_4 $. Ratios for the 6 different geometrical configurations exhibit extremely similar shapes (which mean they are  affected fairly similarly by the free-streaming modeling of the pre-thermalization stage) so results are only shown for configurations 2 and 6. In the approach we use, the free-streaming velocity is  equal to $c$. In case it is smaller, the system would have a slower expansion, reducing
the effects of the pre-thermalization stage.
An interacting model will introduce correlations on the system that will also attenuate  the effects of the pre-thermalization stage. In this way, the magnitude of the effects shown here may be understood as an upper bound. For the subsequent hydrodynamic phase, reducing the free-streaming velocity is important for mitigating the artificial increase of the bulk pressure at the start of the expansion \cite{daSilva:2022xwu}.

We now concentrate on effects of varying the geometry, for a fixed free-streaming time. 
Results are shown for a given nuclear configuration (numbered from 1 to 5, with parameters from Table \ref{tab:nuclear_param}), relative to configuration 6 (expected to be more alike a Zr nucleus). For $\tau_{FS} = 0.4$ fm/$c$,  changes in the nuclear radius and eccentricities compared to their initial values are just a few percent (cf. Fig. \ref{fig:icfs}) and ratios for a given configuration 1--5 with respect to 6 are essentially equal to the ones obtained at initial condition level, so not shown. Unless otherwise stated, we ran 10 million events minimum-bias for each nuclear configuration present in Table \ref{tab:nuclear_param} to obtain the results shown.

\begin{figure}[hbtp]
\begin{center}
  \includegraphics[width=.99\linewidth]{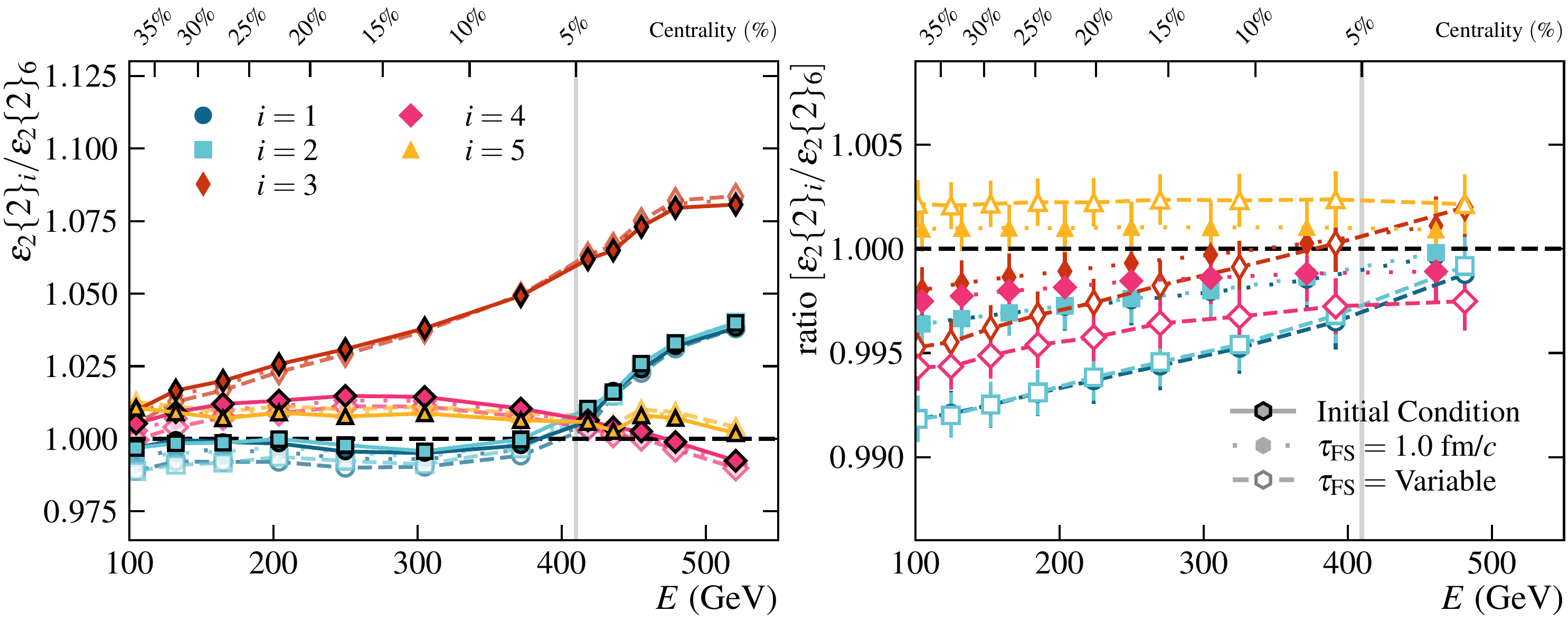}
 \includegraphics[width=.99\linewidth]{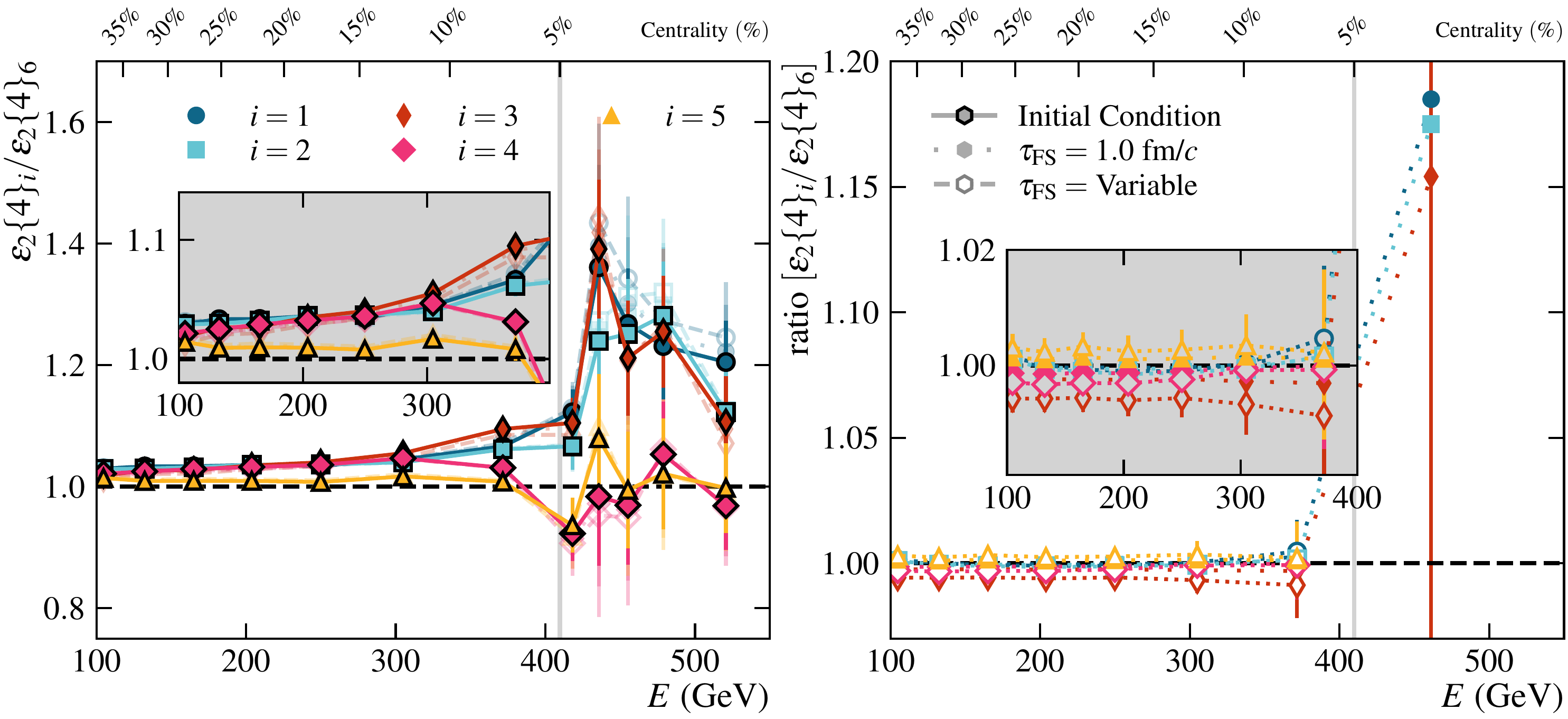}
   \caption{Left panels: Ratios between ellipticity computed for configurations 1 through 5 and configuration 6, as function of total initial energy and with different free-streaming times. From top to bottom, the calculation is made for mean ellipticity, second and fourth order cumulants. Same line patterns as in Fig.~\ref{fig:icfs} are used, with the addition of solid line to represent quantities computed for the initial conditions. Right panels: ratios between the curves of the left hand panel with free-streaming time and the ones for the initial conditions.}
\label{fig:eps2}
\end{center}
\end{figure}

Figure \ref{fig:eps2} illustrates two  effects: the first column represents the influence of distinct nuclear geometries, while the second column emphasizes the effect of free-streaming. The first column of Figure \ref{fig:eps2} shows ratios for the mean ellipticity employing the different nuclear configurations considered in this work. Configurations 1 and 2 have very similar ellipticity and differ only due to triaxiality $\gamma$ which should not affect this quantity. Configuration 3 shows a higher ellipticity than configuration 2 even though both configurations only differ by the inclusion of an octupole deformation $\beta_3$, indicating the non-trivial interplay between $\beta_2$ and $\beta_3$ discussed e.g.~in \cite{Jia:2021tzt}. The decrease in ellipticity from configuration 3 to 4 is expected due to the decrease of the quadrupole deformation. Configurations 4 and 5 have similar ellipticity, with ratios close to 1, indicating that this quantity is insensitive to nuclear radius and diffuseness. The differences in geometry  are more apparent for the most central collisions.

The ratio $\langle\varepsilon_2\rangle_i/\langle\varepsilon_2\rangle_6$ is very close to
the ratio $\varepsilon_2\{2\}_i/\varepsilon_2\{2\}_6$ so only the latter is shown.
This is because
 $\varepsilon_2\{2\}_i/\varepsilon_2\{2\}_6 = \langle\varepsilon_2\rangle_i/\langle\varepsilon_2\rangle_6 \times 
 \sqrt{1+(\sigma^2_i/\langle\varepsilon_2\rangle^2_i)}^2/\sqrt{1+(\sigma^2_6/\langle\varepsilon_2\rangle^2_6)^2}$ and we have checked that the quantities $\sigma^2_n/\langle\varepsilon_2\rangle^2_n$ are nearly equal for $i=1-6$.
 The ratios of $\varepsilon_2\{4\}$ are also shown, even though the error bars are large, because it has been argued that the comparison of $v_2\{2\}$ and $v_2\{4\}$ or $\varepsilon_2\{2\}$ and $\varepsilon_2\{4\}$ for isobars might yield information on nuclear structure \cite{Jia_2023}. The inset illustrates the effect of changing the diffuseness $a$.

Lastly, the free-streaming time has no significant effect on  these ratios of eccentricities. This is indicated more clearly by the right column of Figure \ref{fig:eps2}: eccentricies of a configuration $i=1-5$ relative to that of configuration 6 initially and after free-streaming are changed by at most 1$\%$, and less in central collisions.

\begin{figure}[hbtp]
\begin{center}
\includegraphics[width=.99\linewidth]{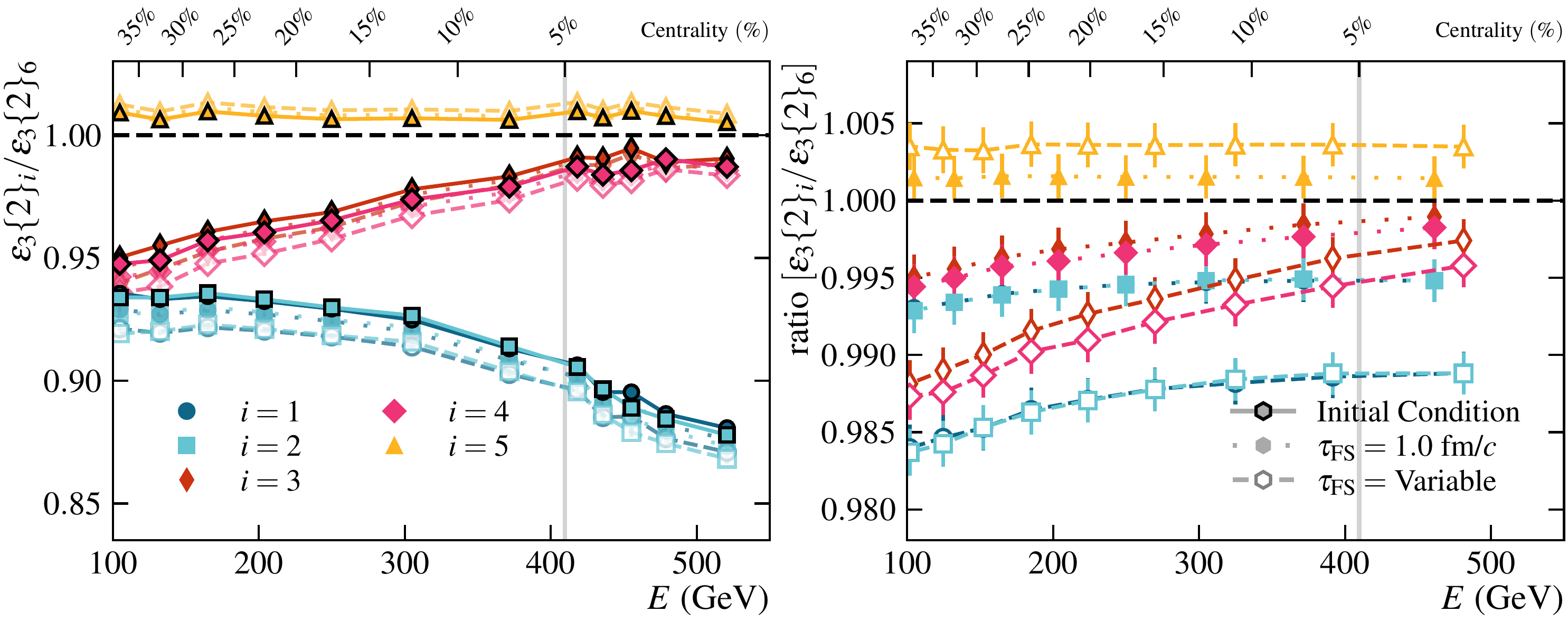}
   \caption{Ratios of triangularities, similar to Fig.~\ref{fig:eps2}.}
\label{fig:eps3}
  \end{center}
\end{figure}

Figure \ref{fig:eps3} shows the triangularity as a function of the total initial energy $E$ for different choices of the free-streaming times. On the left, by comparing the different nuclear configurations it is possible to verify that this quantity is mainly affected by the octupole deformation $\beta_3$ (as expected) and the  diffuseness. The effect is more prominent in central collisions, where the nucleus overlap more and their nuclear geometry becomes increasingly important. Other details such as the nuclear radius, triaxiality, and quadrupole deformation do not affect this quantity. From the right hand side, it can be seen that free-streaming has larger effects for triangularity than for ellipticity for configurations 1 and 2  (divided by that of configuration 6): there is a larger splitting for the two different free-streaming times, almost 1\% for most centralities, which is comparable with the error bars in experimental triangular flow ratios. For configurations 3--5, the effect is a little less pronounced. In any case, the modeling and duration of the pre-thermalization stage may play a role in a very precise determination of the octopole deformation.

\begin{figure}[hbtp]
\begin{center}
\includegraphics[width=.99\linewidth]{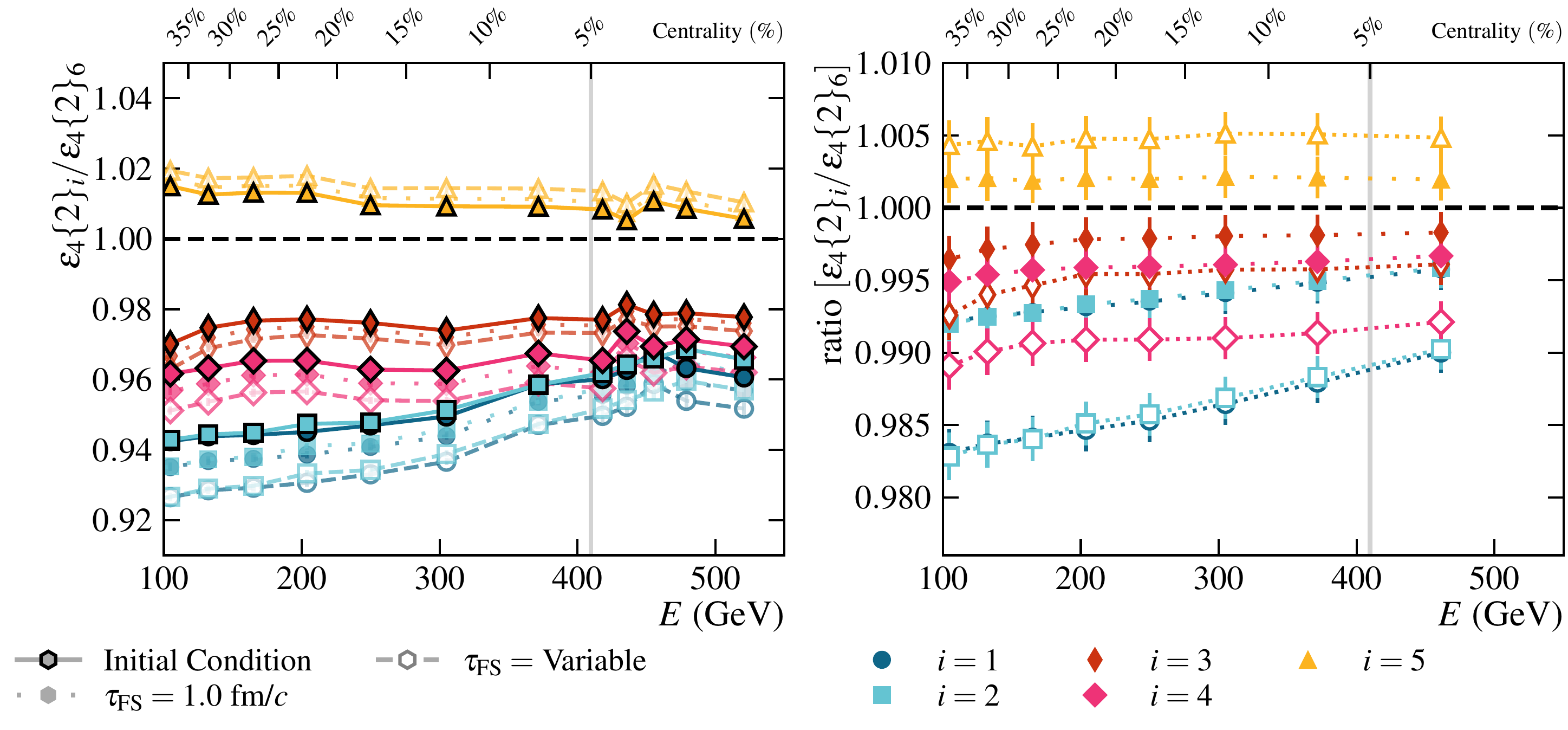}
   \caption{Ratios of higher order eccentricities, similar to Fig.~\ref{fig:eps2}.}
   \label{fig:eps4}
  \end{center}
\end{figure}

Figure \ref{fig:eps4} shows a higher-order eccentricity, $\varepsilon_4$, as function of the total initial energy $E$ for different free-streaming times. When going from configuration 4 to configuration 5, for central collisions, the dependence on the diffuseness parameter is 3\% for central collisions and larger for non central ones. In line with results obtained for triangularity, a progressively larger splitting between the results with different free-streaming times is observed. 
 
\begin{figure}[hbtp]
\begin{center}
\includegraphics[width=.99\linewidth]{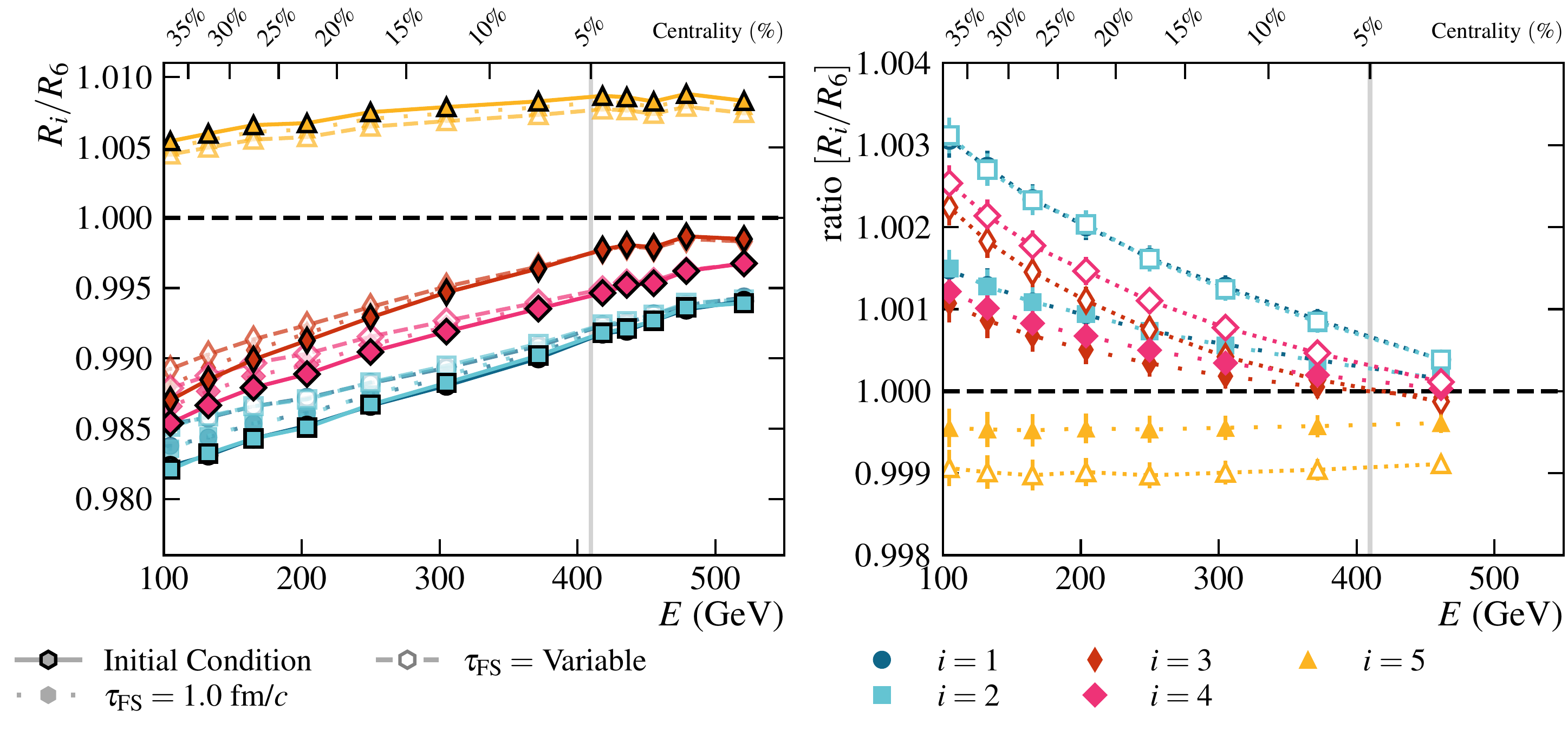}
   \caption{Ratios of root mean square of radius, in a similar fashion as done in Fig.~\ref{fig:eps2} for ellipticity.}
   \label{fig:R}
\end{center}
\end{figure}

The radius of the produced system must increase as it expands over time. The effect of this increase for different $\tau_{FS}$ and nuclear geometries is shown in Figure \ref{fig:R}. Albeit there is a distinct ordering from configuration 1 to configuration 5 in the left panel, this effect is much smaller than the one observed for eccentricities: the radius varies at most  2$\%$, with the effect being smaller for central collisions, regardless of nuclear shape. A non-zero $\beta_3$ together with a large $\beta_2$ (configuration 3) makes this ratio go towards the unity faster when compared to the configuration where mostly either the quadrupole  or octupole deformation (configuration 1, 2, 4) is present. Also, a decreased diffuseness (configuration 5) increases the ratio to above unit, reflecting the overall bigger system size. The right hand side panel shows that even though there is a dependence on the free streaming  time, it is extremely small, particularly for central collisions which may be more interesting to determine the nuclear geometry.

Given that a good mapping is expected between not just $v_2$, $v_3$ and $\varepsilon_2$, $\varepsilon_3$, but also between the event plane angles and initial state symmetry plane angles $\Phi_2$, $\Phi_3$,  it can be used to study the normalized symmetric cumulant $\varepsilon NSC(2,3)$ \cite{Gardim_2017}, 
\begin{align}
\varepsilon NSC(2,3)=\frac{\langle \varepsilon_2^2\varepsilon_3^2\rangle - \langle \varepsilon_2^2\rangle \langle\varepsilon_3^2\rangle}{\langle \varepsilon_2^2\rangle \langle\varepsilon_3^2\rangle}\,
\label{eq:nsc}
\end{align}
as well as the  two-plane correlator 
\begin{align}
    \langle \cos 6(\Phi_2-\Phi_3)\rangle=\frac{\langle\varepsilon_2^3\varepsilon_3^2
    \cos 6(\Phi_2-\Phi_3)\rangle }
    {\sqrt{\langle\varepsilon_2^6\rangle \langle\varepsilon_3^4\rangle}}\,.
    \label{eq:cos}
\end{align}

\begin{figure}[hbtp]
\begin{center}
 \includegraphics[width=.9\linewidth]{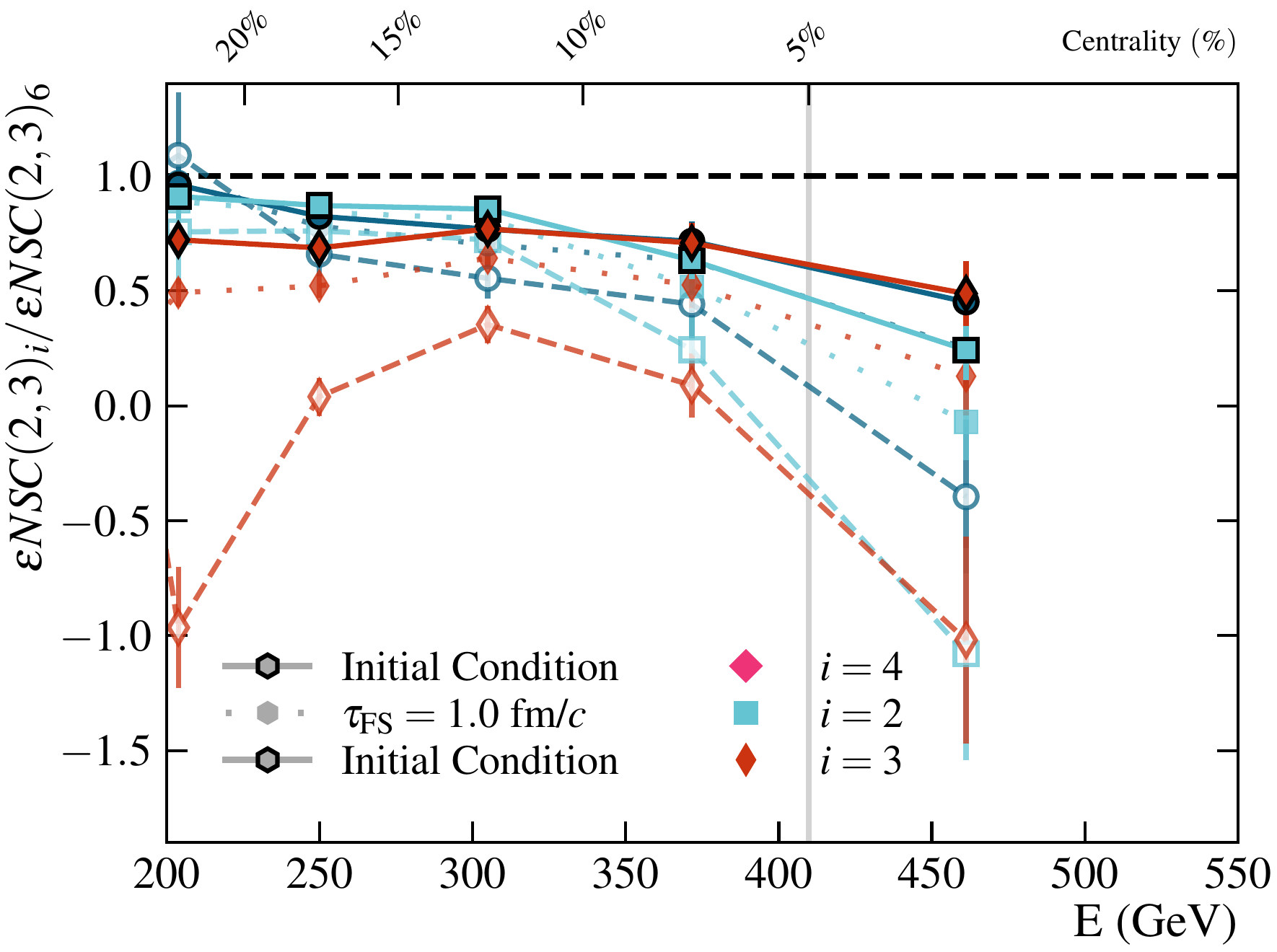}\\
 \includegraphics[width=.9\linewidth]{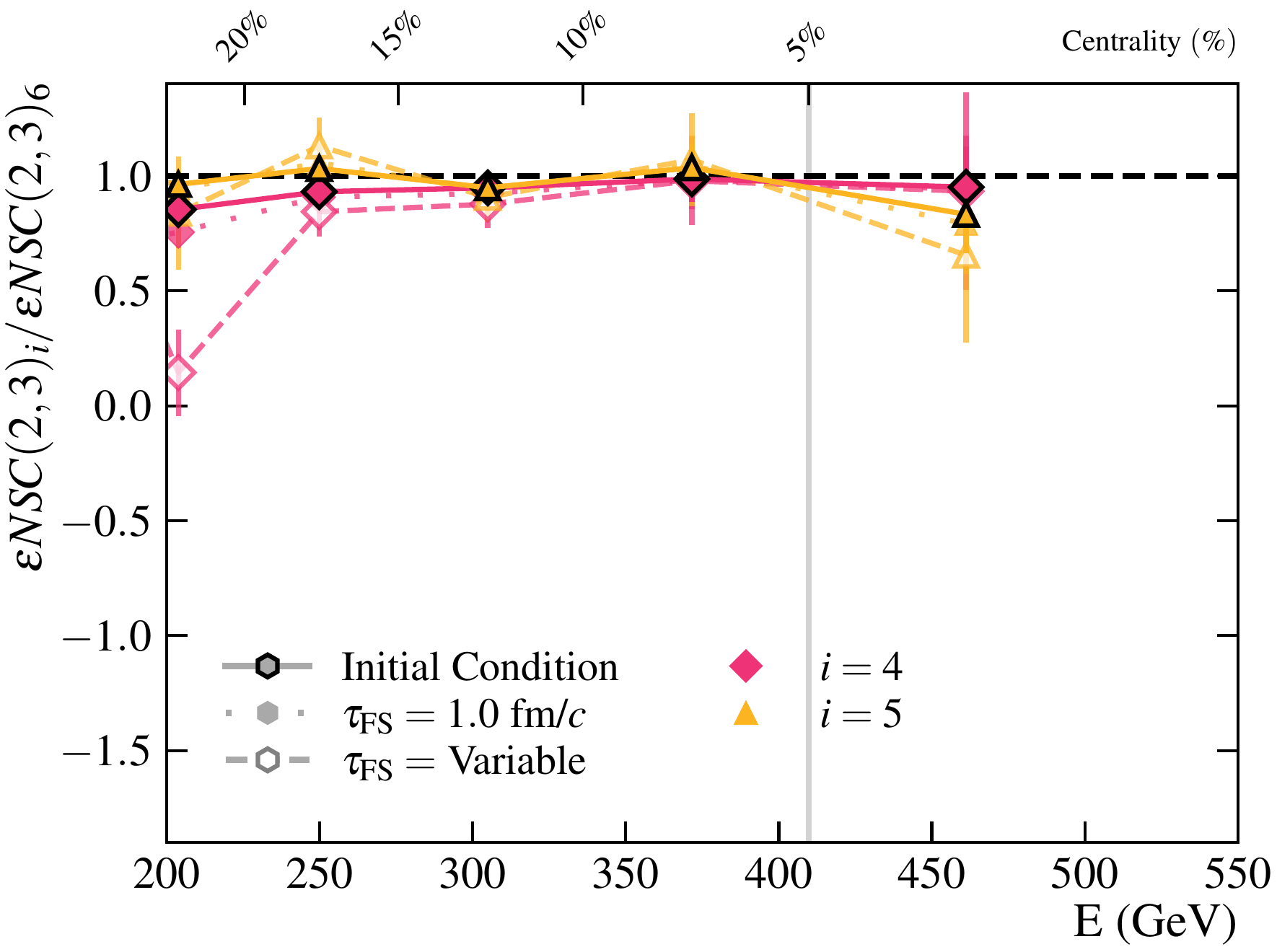}
   \caption{Ratio of normalized symmetric cumulant $\varepsilon NSC(2,3)$ computed for the  geometrical configurations $i=1-5$ compared to $i=6$. }
\label{fig:SC23}
  \end{center}
\end{figure}

In Fig. \ref{fig:SC23},
for the normalized symmetric cumulant computed directly from the initial conditions,
configurations 1, 2, 3 are similar and 4 and 5 are  fairly equal and are shown separately for clarity. Free streaming seems to have a strong effect, and might change once or twice the sign of the ratio, however error bars are large, so more statistics would be necessary to confirm this. There are oscillations for $E<200$ GeV so results are shown only for larger values.
The changes of sign can be traced back to slightly different free-streaming time dependencies for each nuclear geometry considered: as the system evolves in time, the individual quantities $\varepsilon NSC(2,3)$ appearing in this ratio becomes less negative, with each of them crossing zero at different energy values. Configuration 6 only shows a single change of sign (positive to negative) that always happens for values of energy lower than the ones shown in our plots, meanwhile, configurations 2 and 3 for instance, present two clear changes of sign (positive to negative and back to positive), with the second change happening in the most central bin considered. At this point, it is interesting to discuss what would happen if $\alpha$ were negative in the definition of $\tau_{FS}$ in case (iii). In the leftmost bin shown in the figure,
according to fig. \ref{fig:taufs}, the values of $\tau_{FS}$ are fairly equal so the ratio of  $\varepsilon NSC(2,3)$ is the same for $\alpha$ positive or negative. In the rightmost bin,
again from fig. \ref{fig:taufs},
$\tau_{FS}$ would decrease from $\approx 1.48$ fm/c to $\approx 1.44$ fm/c. Therefore the value of the ratio of  $\varepsilon NSC(2,3)$ would be between the value for variable $\tau_{FS}$ and that for $\tau_{FS}=1$ fm/c, but nearest to the former. So the sign changes (top panel) would not be much affected.

The ratio of two-plane correlators, fluctuates around zero and suffer from lack of statistics so is not shown. Another interesting angle correlator for which data have been obtained is $ac2(3)=\langle v_2^2  v_4
    \cos 4(\Psi_2-\Psi_4)\rangle$. Since it cannot be predicted for all centralities with just $\varepsilon_4$ \cite{Gardim:2011xv,Gardim:2014tya}, we leave it to a future study from hydrodynamics.
We note however that it has a good potential to discriminate geometry in central collisions \cite{Zhao_2023}.

The observables presented so far, except perhaps $\varepsilon NSC(2,3)$, exhibit minimal sensitivity to the effects of free-streaming and triaxiality. One observable known to be sensitive to triaxiality is the Pearson correlator between event anisotropic flow and  event mean transverse momentum $[p_T]$ in a given centrality bin~\cite{Bally:2021qys,ATLAS:2022dov}. This correlator has been identified as a strong candidate for assessing the effects on triaxility in isobars \cite{Jia:2021qyu}. Previous studies were limited to $n=2$ and utilized estimators that lack proper correlation with mean transverse momentum: Ref.~\cite{Giacalone:2020dln} showed that a good proxy to estimate  this type of correlator is to replace the mean momentum by the ratio between total energy and total entropy in a very narrow centrality bin and both  $v_2 \propto \varepsilon_2$ and  $v_3 \propto \varepsilon_3$ can be considered. Therefore  we calculate the following Pearson correlator: 
\begin{align}
    \rho_n = \frac{\langle v_n^2 [p_T]\rangle - \langle v_n^2 \rangle \langle [p_T]\rangle}{\sigma_{v_n^2}\sigma_{[p_T]}}
    \approx
    \frac{\langle \varepsilon_n^2 E/S\rangle - \langle \varepsilon_n^2 \rangle \langle E/S\rangle}{\sigma_{\varepsilon_n^2}\sigma_{E/S}} \,,
    \label{eq:rhon}
\end{align}
where the averages are taken in centrality bins of 0.25\% centrality and $\sigma_{\varepsilon_n^2}$ and $\sigma_{E/S}$ are the standard deviation of 
$\varepsilon_n$ and $E/S$ in those centrality bins, respectively.
$S$ is computed as $S=\int d^2 x_\perp s \left(N\left[\frac{T_A(x,y)^p+T_B^p(x,y)}{2}\right]^{1/p}\right)$
where the entropy density $s$ is obtained from the equation of state.

\begin{figure}[hbtp]
\begin{center}
 \includegraphics[width=.9\linewidth]{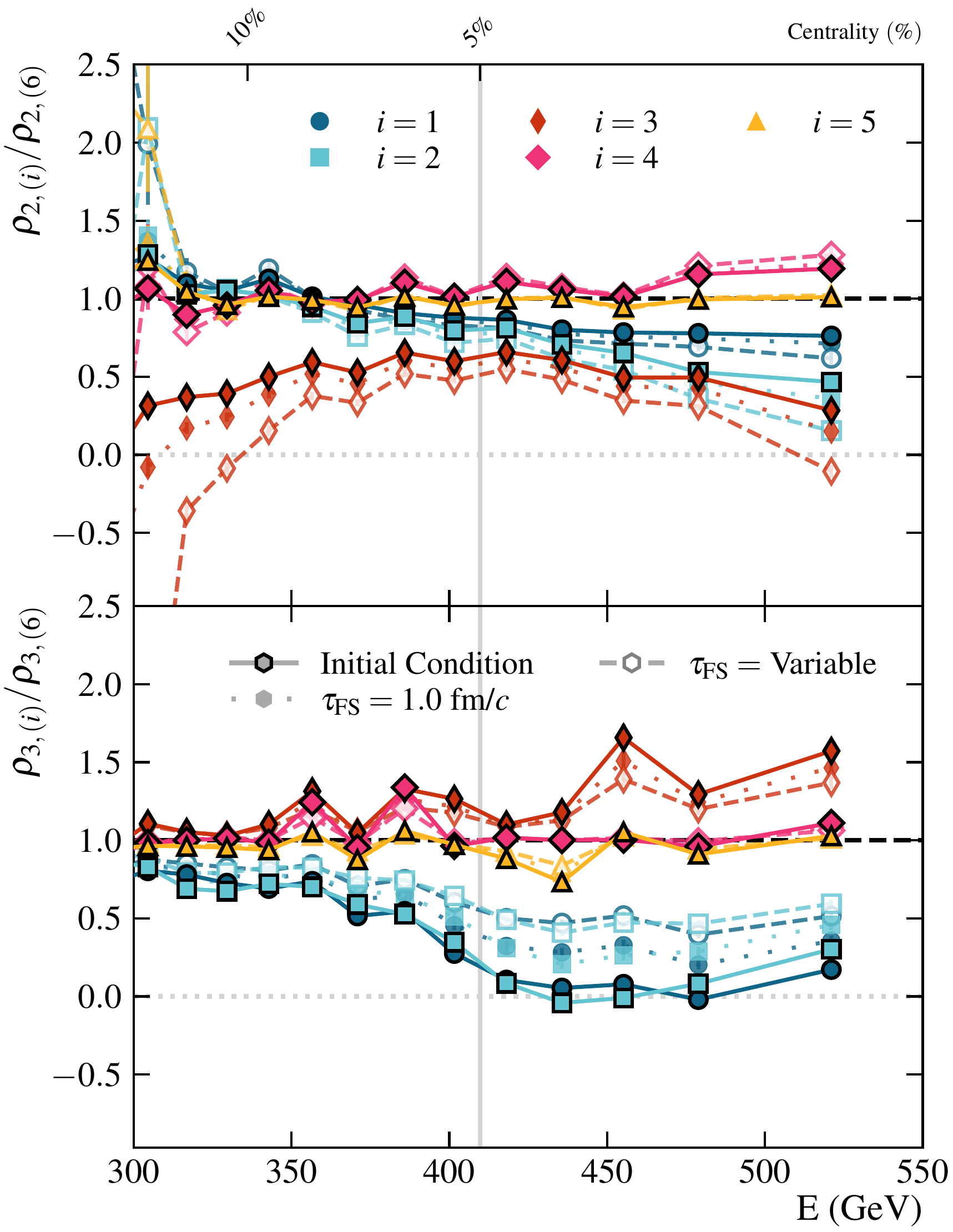}
   \caption{Top panel: Pearson correlation between eccentricity squared and the ratio of total energy by total entropy as function of total initial energy for initial conditions. Lower panel: The same, but replacing ellipticity by triangularity in the $\rho_n$ correlator.}
\label{fig:rhon}
  \end{center}
\end{figure}

Based on the above results, where a mild to none effect of free-streaming stage was observed in $\varepsilon_{2,\,3}$ and having that both $E$ and $S$ are conserved in the present description of the pre-thermalization stage, one might be tempted to not expect any important effect from the free-streaming stage for this quantity. It is possible to verify in Fig.~\ref{fig:rhon}, which presents the ratios of $\rho_{n=2,\,3}$ for configurations 1 through 5 with respect to configuration 6, that is not the case. As other results presented here, these results were obtained with 10 million events for each configuration and, following ~\cite{Giacalone:2020dln}, the calculation was initially done for centrality bins of 0.25\% that were later averaged in groups of 1\% to improve visibility.

Our results confirm that the effect the triaxiality parameter $\gamma$ is particularly clear when comparing configurations 1 and 2 for $\rho_2$ in the top panel of Fig.~\ref{fig:rhon}. We also report that the splitting between the results of these two configurations increases with the free-streaming time, starting from $\approx 0.3$
at the initial condition level and reaching $\approx 0.5$
when considering a centrality dependent free-streaming duration. Comparing configurations 2 and 3, shows that the value of $\beta_3$ has a strong effect, lowering the correlation in the $\approx 5\% - 13\%$ centrality range. Moreover, one verifies that the free-streaming time plays an interesting role, inducing a single or a double sign change depending on how long the system stays in the pre-thermalization stage. While we still lack final experimental results for this correlator, such a signature may eventually help to constrain the duration of the pre-hydrodynamic phase in isobar collisions. Next, reducing the quadrupole deformation while keeping the octupole deformation fixed shifts the correlation to values around unity with a much reduced free-streaming time dependence. For the configuration 5, the result is brought even closer to unity. Using these results together with the ones presented in Fig.~\ref{fig:eps2}, \ref{fig:eps3} indicates that it might be possible to constrain $\beta_2$, $\beta_3$ and $\gamma$ simultaneously from observations of anisotropic flow and the correlation of elliptical flow with mean transverse momentum.

We have checked that using  different estimators for $[p_T]$ (taken from ~\cite{Giacalone:2020dln}), the effect of the free-streaming time is still substantial.

Moving to $\rho_3$, and assuming that the estimator (\ref{eq:rhon}) is still a good proxy for the final state observable (as is the case for high-energy Pb+Pb collisions~\cite{Giacalone:2020dln}), one sees that while the dependence on the triaxiality parameter becomes weaker, the effect of free-streaming time is still visible for configurations 1--3. In addition, adding an octupole deformation on top of a quadrupole deformation (configuration 3) shifts the correlation to even larger values in central collisions, providing a clear separation with configurations 1 and 2  where only the quadrupole deformation is present. Configurations 4,  5 and 6 all have similar $\rho_3$, indicating that this observable is independent of diffuseness and distribution radius  in the centrality range considered.

\begin{figure}[hbtp]
\begin{center}
 \includegraphics[width=.9\linewidth]{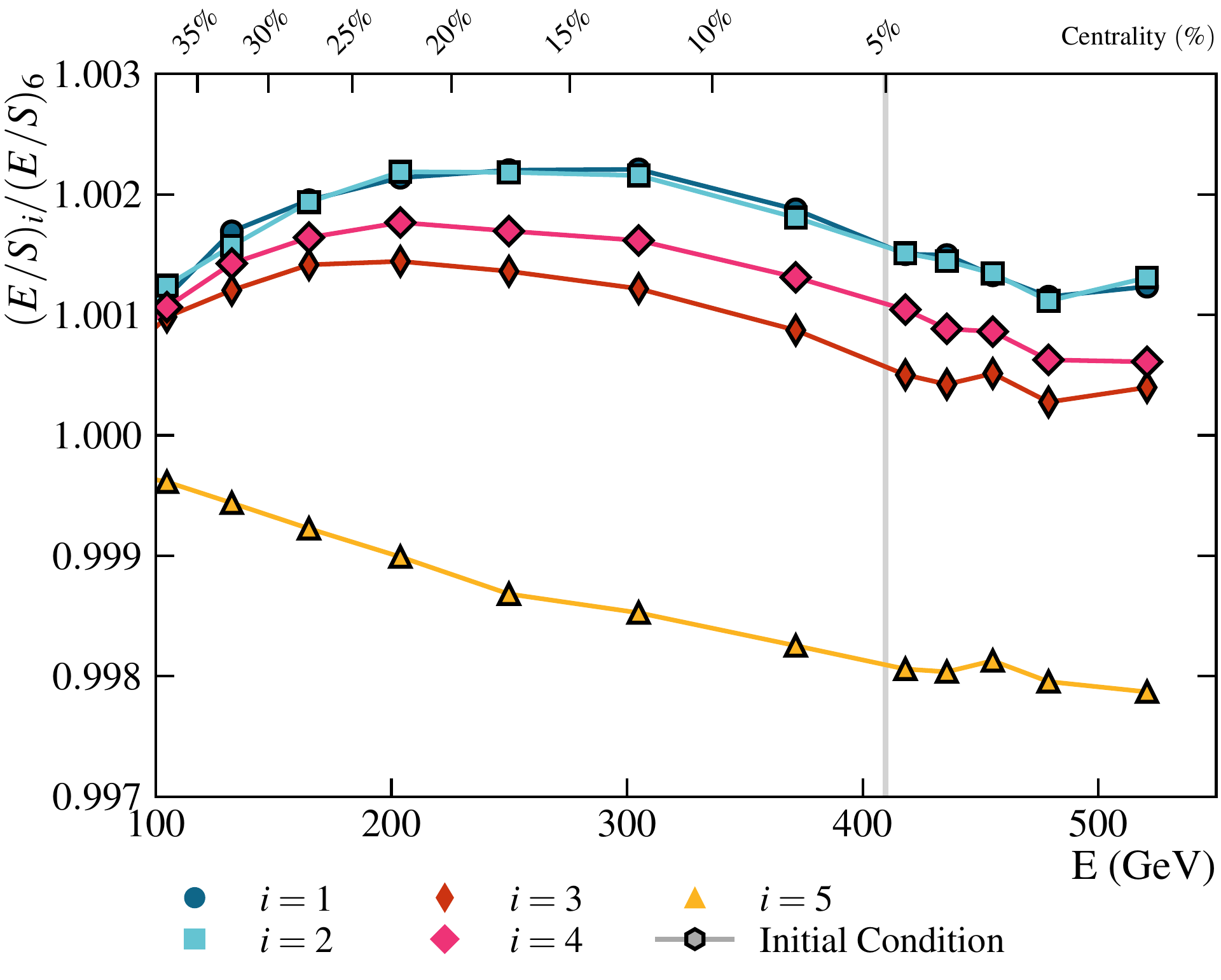}
   \caption{Expected value for the distribution of $E/S$ (in a centrality bin) for configuration $i$ divided by that of configuration 6 as function of centrality.}
\label{fig:pt}
  \end{center}
 \end{figure}

The $\alpha<0$ case can be studied in a similar way as done for Fig. \ref{fig:SC23}, one sees that it would not alter the general behavior of $\rho_2$ and $\rho_3$.

\begin{figure}[hbtp]
\begin{center}
 \includegraphics[width=.9\linewidth]{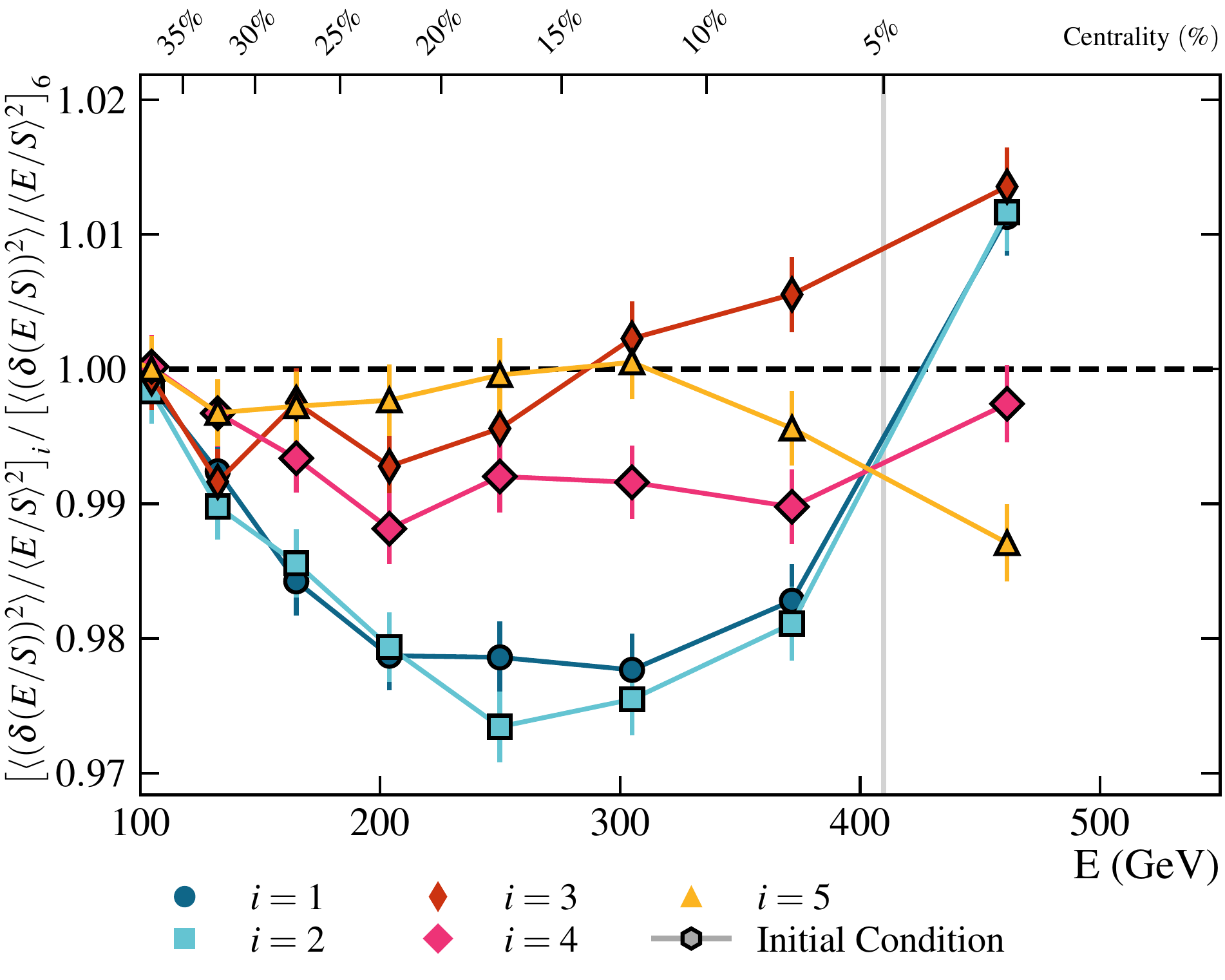}
   \caption{Same as Fig.~\ref{fig:pt} for the variance.}
\label{fig:pt2}
  \end{center}
 \end{figure}

\begin{figure}[hbtp]
\begin{center}
  \includegraphics[width=.9\linewidth]{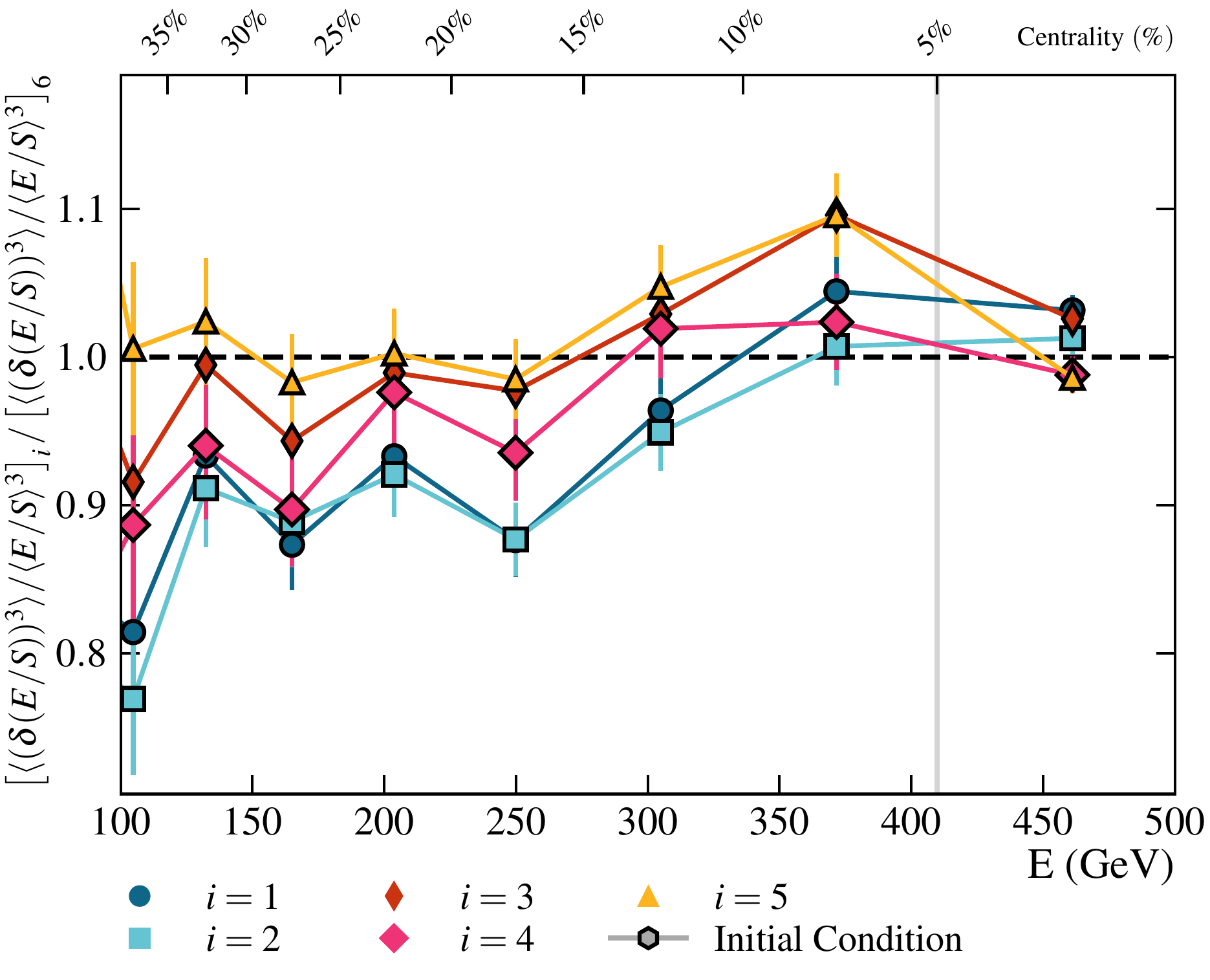}
   \caption{Same as Fig.~\ref{fig:pt} for the third cumulant.}
\label{fig:pt3}
   \end{center}
 \end{figure}

Another possibility to determine the geometry is the study of the distribution of event mean transverse momentum  in a given centrality bin, $[p_T]$. As argued above, a good estimator for $[p_T]$ is $E/S$. It is expected  that a good estimator for $\langle \delta [p_T]^p \rangle/ \langle [p_T]\rangle^p$ would be $\langle (\delta (E/S))^p\rangle / \langle E/S\rangle^p$ where $\delta O\equiv O-\langle O \rangle$ (as far as we know this has not been tested throughout). The ratio $\langle E/S\rangle_i/\langle E/S \rangle_6$ as function of centrality is shown in Fig.~\ref{fig:pt}. It is computed directly from the initial conditions (as E/S is conserved) and exhibit a clear ordering: configurations $i=1-2$ have the highest values for the ratio, above 1, and do not depend on the triaxiality. Configuration 3 is also  above 1  but introducing $\beta_3$ has decreased the ratio. For configuration 4, decreasing  $\beta_2$ increases the ratio. For configuration 5, increasing $a$ decreases the ratio.  However, all these effects are quite small. Fig.~\ref{fig:pt2} is more interesting: the ratios 
$\left[\langle (\delta (E/S) )^2\rangle/ \langle E/S\rangle^2\right]_i/\left[\langle (\delta (E/S))^2\rangle/ \langle E/S\rangle^2\right]_6$ are shown and larger effects are observed. Configurations $i=1-2$ have the most extreme values for the ratio, below 0.98 and above 1.01, and do not depend on the triaxiality. Configuration 3  also  reaches above 1.01  but introducing $\beta_3$ has pushed the ratio towards 1 in general (by about 2\%). For configuration 4, decreasing  $\beta_2$ decreases the ratio. For configuration 5, increasing $a$ pushes  the ratio very near  1 except for very central collisions. This quantity might be useful to further constrain the value of $\beta_2$ given that curves with small values of this parameter are well separated from those with higher values for central collisions. Finally in Fig.~\ref{fig:pt3}, $\left[\langle (\delta (E/S) )^3\rangle/ \langle E/S\rangle^3\right]_i/\left[\langle (\delta (E/S) )^3\rangle/ \langle E/S\rangle^3\right]_6$ is shown. It has been claimed \cite{Jia:2021qyu} that this quantity could also be used to constrain $\gamma$, but the effect seems very small. Though the error bars are big, some geometrical effects can be seen, for example, introducing $\beta_3$ increases the ratio and diminishing $\beta_2$ decreases it.

In \cite{Jia:2021qyu}, some quantities comparable to those we computed have be obtained as well but 
with a  less efficient estimator for $[p_T]$. As a general rule, the effect of geometry is qualitatively similar.

\section{Conclusion}

We investigated the effects of a pre-thermalization stage in isobar collision systems (namely \Ru+\Ru and \Zr+\Zr), using free-streaming as an extreme case. The main effects anticipated were observed: the smearing out of small-scale features and reduction of spatial anisotropies (which get transferred to momentum distribution anisotropy). Despite the absolute magnitude of these effects being large, it is not the same when doing ratios for a given $\tau_{FS}$. In this case, for ratios of ellipticity and rms radii, differences are less than 1\% and  unlikely to be relevant when comparing to data.  Ratios of higher order eccentricities, however, show differences that can be significant when comparing with precision measurements such as the ones recorded for isobar collisions. Within our statistics, not much can be learned from  the two-plane correlator for elliptic and triangular flows but interestingly the normalized symmetric cumulant $\varepsilon NSC(2,3)$  seems sensitive to the duration of free-streaming.
These studies of eccentricities yield information on the parameters $\beta_2$, $\beta_3$ and $a$.

Furthermore, we studied the effects that nuclear geometry could have on the correlation between anisotropic flow and mean momentum to get information on $\gamma$. In \cite{Giacalone:2020dln}, the estimator $\rho(\varepsilon_2^2,E/S)$ was shown to better reproduce the hydrodynamic correlator $\rho(v_2^2,[p_T])$. Here  for the first time,  it is used to study isobar collisions. A several percent difference between the configurations $i=1$ and 2 (with and without triaxiality) is indeed observed for $\rho_2$. In addition, and this is the main result as far as free-streaming effects are concerned, the duration of the pre-thermalization stage, regulated by $\tau_{FS}$, has a noticeable and diverse effect on these ratios of correlators depending on the description of the nuclear geometry: for $\rho_2$, it is able to enhance the separation between results with and without triaxiality effects (configurations 1 and 2); for $\rho_3$ it generates a non-zero correlation between $\varepsilon_3$ and $E/S$ (proxies for $v_3$ and the mean transverse momentum, respectively) for the same configurations. Another distinct effect for $\rho_2$ appears for configuration 3 (where in addition to the quadrupole deformation of configurations 1 and 2, an octupole deformation is introduced), a double sign change  for the largest free-streaming times ($1.35 \lesssim \tau_{FS}\, ({\rm fm/c}) \lesssim 1.49$) appears, which can be contrasted to no or perhaps one sign change for initial condition  and  free-streaming time equal to $1$ fm/c. If verified experimentally, these features provide another tests of the predictive power of state-of-the-art hybrid hydrodynamic models~\cite{Giannini:2022bkn}.
 
The analysis with $E/S$ as estimator for $[p_T]$ is also used to study the distribution of  $[p_T]$ and its first momenta, for initial conditions only (because $E/S$ is conserved during free-streaming) for the first time. The ratio of variances are most promising to constrain $\beta_2$  in addition to the eccentricities analysis. The ratio of third order cumulant which in principle could yield information on $\gamma$ as well, does not seem to lead to a strong effect. 

In a future work, we will test how good the estimators presented here are, by comparing with actual hydrodynamic simulations. In turn, this will permit to select the best estimators and  use them to perform a bayesian analysis to extract nuclear geometry.

\section{Acknowledgements}

F.G.G. was supported by Conselho Nacional de Desenvolvimento Cient\'{\i}fico  e  Tecnol\'ogico  (CNPq grant 306762/2021-8). K.P.P. and W.M.S. acknowledge  support from Funda\c{c}\~ao de Amparo \`a Pesquisa do Estado de S\~ao Paulo (respectively via grants 2020/15893-4 and 2021/01670-6, 2022/11842-1). All authors acknowledge support from
Funda\c{c}\~ao de Amparo \`a Pesquisa do Estado de S\~ao Paulo (FAPESP grant 2018/24720-6) and project INCT-FNA Proc.~No.~464898/2014-5. A.V.G. has been partially supported by CNPq. The authors thank the HPC resources provided by Information Technology Superintendence (HPC-STI) of University of São Paulo (Aguia cluster). 
The authors acknowledge the National Laboratory for Scientific Computing (LNCC/MCTI, Brazil), through project COL-ISO as well as 
the ambassador program (UFGD), subproject FCNAE, which providing HPC resources from the SDumont supercomputer 
URL: \url{http://sdumont.lncc.br}.
The authors thank the coordinators of the EMMI task force "Nuclear physics confronts relativistic collisions of isobars",    
G.Giacalone, J. Jia, V. Som\`a, Y. Zhou,  who first suggested this line of research.

\bibliographystyle{apsrev}
\bibliography{main}

\end{document}